\def\year{2022}\relax
\documentclass[letterpaper]{article} 
\usepackage{aaai22}  
\usepackage{times}  
\usepackage{helvet}  
\usepackage{courier}  
\usepackage[hyphens]{url}  
\usepackage{graphicx} 
\usepackage{natbib}  
\usepackage{caption} 
\DeclareCaptionStyle{ruled}{labelfont=normalfont,labelsep=colon,strut=off} 
\usepackage{algorithm}
\usepackage{algorithmic}

\usepackage{newfloat}
\usepackage{listings}
\floatstyle{ruled}
\newfloat{listing}{tb}{lst}{}
\floatname{listing}{Listing}
\usepackage{mathtools}

\title{3D-OOCS: Learning Prostate Segmentation with Inductive Bias}
\author {
    Shrajan Bhandary\textsuperscript{\rm 1},
     Zahra Babaiee\textsuperscript{\rm 1},
     Dejan Kostyszyn\textsuperscript{\rm 2},
     Tobias Fechter\textsuperscript{\rm 2},
     Constantinos Zamboglou\textsuperscript{\rm 2},
     Anca-Ligia Grosu\textsuperscript{\rm 2},
     Radu Grosu\textsuperscript{\rm 1}
}
\affiliations {
    \textsuperscript{\rm 1} Technische Universität Wien, Vienna, Austria\\
    \textsuperscript{\rm 2} Department of Radiation Oncology, University Medical Centre Freiburg, Germany \\
    shrajan.bhandary@tuwien.ac.at, zahra.babaiee@tuwien.ac.at,
    dejan.kostyszyn@uniklinik-freiburg.de, 
    tobias.fechter@uniklinik-freiburg.de, 
    constantinos.zamboglou@uniklinik-freiburg.de, 
    anca.grosu@uniklinik-freiburg.de, 
    radu.grosu@tuwien.ac.at
}

\usepackage{bibentry}

\begin{document}

\maketitle

\begin{abstract}

Despite the great success of convolutional neural networks (CNN) in 3D medical image segmentation tasks, the methods currently in use are still not robust enough to the variety of image properties or artefacts. To this end, we introduce OOCS-enhanced networks, a novel architecture inspired by the innate nature of visual processing in the vertebrates. With different 3D U-Net variants as the base, we add two 3D residual components to the second encoder blocks: on and off center-surround (OOCS). They generalise the ganglion pathways in the retina to a 3D setting. The use of 2D-OOCS in any standard CNN network complements the feedforward framework with sharp edge-detection inductive biases. The use of 3D-OOCS also helps 3D U-Nets to scrutinise and delineate anatomical structures present in 3D images with increased accuracy. We compared the state-of-the-art 3D U-Nets with their 3D-OOCS extensions and showed the superior accuracy and robustness of the latter in automatic prostate segmentation from 3D Magnetic Resonance Images (MRIs). 
\end{abstract}

\section{Introduction}
Prostate cancer (PCa) is one of the leading causes of death in males. Radiation therapy (RT) is one of the main options of medical care, but as with most treatments, it has its drawbacks. Moreover, due to the unique traits of different patients, there is an urgent need to personalize and improve it. Personalization can be achieved by considering the individual PCa biology and the disease process of each patient. Quality improvement can be accomplished by increasing the robustness of prostate segmentation, in addition to that of intraprostatic gross-tumour-volume (GTV) segmentation~\cite{thno61207}. Both are important for a variety of applications, such as prostate-volume measurement, targeted biopsies, pre-radiation treatment planning, and monitoring disease progression over time~\cite{goldenberg_NATURE_2019}. However, manual segmentation of structures in the pelvic region by physicians, based on medical images, underlies significant interobserver heterogeneity and is time consuming~\cite{steenbergen_2015, rischke_2013}.

\begin{figure}[t]
\centering
\includegraphics[width=0.35\textwidth]{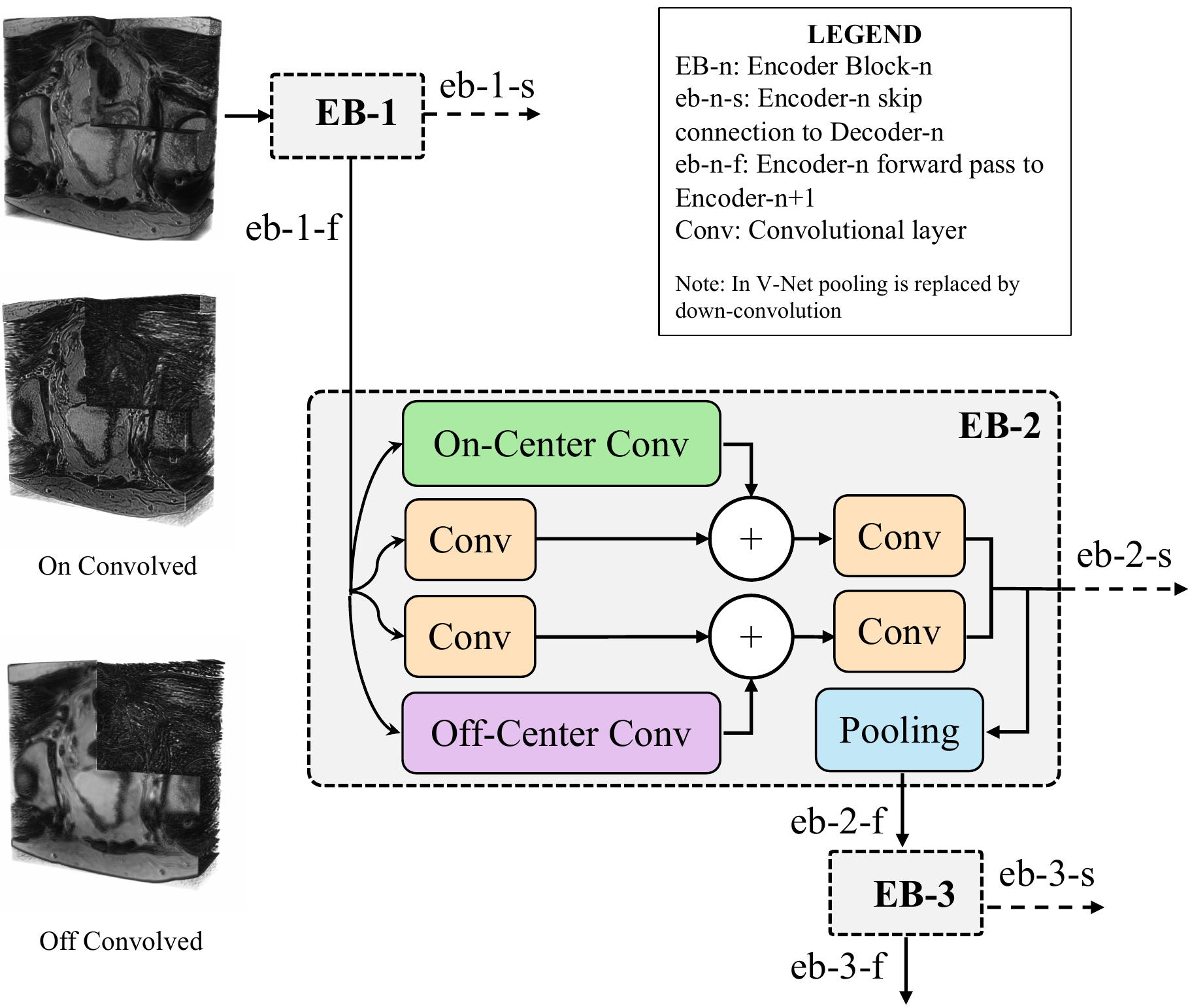}
\caption{Extending U-Net-like architectures with the two 3D-OOCS inductive biases enhances their segmentation performance and their robustness to distribution shifts.}
\label{fig:oocsArchitecture}
\end{figure}

Since the success of AlexNet~\cite{alexnet_2012}, the biomedical-imaging community has investigated various CNN architectures as promising solutions to enhancing cancer-detection efficiency~\cite{3d_survey_singh_2020}. U-Nets are a particular type of CNNs that were originally designed to work with small datasets and yield a highly accurate segmentation~\cite{10.1007/978-3-319-24574-4_28}, first in 2D and later on in 3D medical imaging scenarios~\cite{cicek_3d_2016}. The U-Net network has an encoder-decoder structure, where the encoder performs as a classification network based on annotated images. In contrast, the decoder is (loosely) a dual network that reconstructs the image from its discriminative features (low-dimensional feature space) in the original pixel-level space (high-dimensional features). 

Inspired by the OOCS retinal circuitry and motivated by the significant efficacy of 2D OOCS in image-classification tasks~\cite{pmlr-v139-babaiee21a}, we investigated if the 3D OOCS extension can improve the accuracy and robustness of U-Net-like architectures. We first designed the 3D OOCS kernels and then modified the second encoder block of our U-Net baselines, with On and Off parallel pathways. Our experiments on prostate segmentation show that introducing OOCS blocks to a U-Net-like network not only improves its performance but also increases the network's robustness to test-set distribution shifts, such as noise, blurring, and motion.

\section{Related Work}
\label{related_work}

\subsection{Biologically Inspired Architectures}

\subsubsection{Vision models.} Originally, they were inspired by neuro\-science and psychology. In the meantime, both neuroscience and artificial intelligence (AI), in particular computer vision, have enormously progressed. Nevertheless, most of today's neural networks remained only loosely inspired by the visual system. The interaction between the fields has become less common, compared to the early age of AI, despite the importance of neuroscience in generating accelerating ideas for AI~\cite{hassibis, HASSABIS2017245}.

\subsubsection{2D OOCS kernels.}
A recent paper introduced OOCS parallel pathways to CNNs and evaluated their effect in several image-classification and robustness scenarios~\cite{pmlr-v139-babaiee21a}. OOCS-CNNs outperformed the CNNs baselines in accuracy and surpassed state-of-the-art regularisation methods in unseen lighting conditions and noisy test sets~\cite{pmlr-v139-babaiee21a}. However, that paper only studied OOCS for image classification tasks and was limited to 2-dimensional natural image datasets. In this paper, we investigate 3D OOCS, which are beyond what exists in the human visual system. We also show the superior performance of networks incorporating 3D OOCS compared to their U-Net-like baselines in the 3D prostate segmentation task. Our experiments show that similar to the 2D version, 3D OOCS extensions significantly enhance accuracy and robustness.

\section{Method}
In this section we first show how we compute the 3D OOCS kernels. We then discuss how to extend a U-Net-like architecture with OOCS inductive biases.

\subsection{3D OOCS Kernels}
OOCS receptive fields are captured with a difference of two Gaussians (DoG)~\cite{rodieck_1965}. For 2D kernels, Equation~\eqref{eq1:petrov} computes their center and surround weights~\cite{Petkov2005ModificationsOC,kruizinga_petkov_2000}:
\begin{equation}
\label{eq1:petrov}
DoG_{\sigma,\gamma}(x,y) =  \frac{A_c}{\gamma^2}\, e^{-\frac{x^2+y^2}{2\gamma^2\sigma^2}} - A_s\,e^{-\frac{x^2+y^2}{2\sigma^2}}
\end{equation}
In Equation~\eqref{eq1:petrov}, $\gamma$ is the ratio of the radius of the center to the surround, $\gamma < 1$, $\sigma$ is the variance of the Gaussian function, and $A_c$ and $A_s$ are the center and surround coefficients.

To construct 3D-OOCS kernels, the naive way would be to use Equation~\eqref{eq1:petrov}, and repeat the kernels in the third axis. However, this approach would result in a cylindrical center-surround structure, instead of a spherical one, with excitatory center, and an inhibitory surround. We therefore use a 3D DoG, with three parameters, extending the 2D version: 
\begin{equation}
\label{eq:petrov_3d}
DoG_{\sigma,\gamma}(x,y,z) =  \frac{A_c}{\gamma^3}\, e^{-\frac{x^2+y^2+z^2}{2\gamma^2\sigma^2}} - A_s\,e^{-\frac{x^2+y^2+z^2}{2\sigma^2}}
\end{equation}

We set the absolute value of the sum of the negative and positive weights, to be equal to an arbitrary value $c\geq1$. This ensures the proper balance between excitation and inhibition and the kernel weights will be large enough.
\begin{equation}
\iiint [DoG_{\sigma,\gamma}(x,y,z)]^{+} dx dy dz = c,
\end{equation}
\begin{equation}
\iiint [DoG_{\sigma,\gamma}(x,y,z)]^{-} dx dy dz = -c
\end{equation}

By setting the sum of negative and positive values to be equal, the coefficients $A_c$ and $A_s$ will be equal in the infinite continuous case. To show this, we can write the DoG equation in polar coordinates: 
\begin{equation}
\int_{0}^{r_s}\!\!\!\!\int_{0}^{\pi}\!\!\!\!\int_{0}^{2\pi}\!\!\!\!\!\! r^2\,sin(\varphi)\,(\frac{A_c}{\gamma^3}\ e^{-\frac{r^2}{2\gamma^2\sigma^2}} - A_{s}\,e^{-\frac{r^2}{2\sigma^2}}) d\theta d\varphi dr
\end{equation}
Here $r$ is the radius of the central sphere. We can calculate the integral when $r\,{\to}\,\infty$, and by knowing the sum of positive and negative weights equals to zero, we have:

\begin{equation}
\label{eq:petrov}
\frac{2\sqrt{2}\pi^{\frac{3}{2}}}{(\frac{1}{\sigma^2})^{\frac{3}{2}}} (A_c - A_s) = 0 \ \ \Rightarrow\ \  A_c = A_s
\end{equation}

Assuming that the values of the coefficients are still close enough in the finite discrete case as well, we can approximate the variance as follows:
\begin{equation}
\label{eq:sigma}
\sigma \approx \frac{r}{\gamma}\sqrt{\frac{1-\gamma^2}{-6\ln{\gamma}}}
\end{equation}

\subsection{3D OOCS Encoder Blocks}
We use Equation~\eqref{eq:petrov_3d} to calculate the kernel weights for the On and Off (same equation, but with signs inverted) convolutions. Note that we only need to determine the kernel size, ratio of the radius of the center to surround, and sum of positive and negative values to compute the fixed kernels. Here we set $\gamma=\frac{2}{3}$ and $c=3$.

For a given input  $\chi$, we calculate the On and Off responses by convolving $\chi$ with the computed kernels separately:
\begin{equation}
\begin{array}{c@{\ }c@{\ }l}
\chi_{\rm On}[x,y,z] &=& (\chi * +DoG[r,\gamma, c])[x,y,z]\\[1mm]
\chi_{\rm Off}[x,y,z] &=& (\chi * -DoG[r,\gamma, c])[x,y,z]
\end{array}
\end{equation}

We extend a 3D U-Net-like segmentation network to a 3D OOCS network, by changing one of its primary encoder blocks, to a 3D OOCS encoder block, as shown in Figure~\ref{fig:oocsArchitecture}. To this end, we first divide the convolution layers of the block into two parallel pathways, with half of the filters of the convolutions in the original layer, in each divided convolution, imitating the so-called On and Off pathways in the retina. We add the 3D On and Off convolved inputs of the block to the activation maps of the first convolution layers of the pathways. Finally, we concatenate the activation maps of the last layers on the pathways, resulting in the outputs with the same shapes as the outputs of the original block. 

\section{Experimental Evaluation}
The proposed 3D OOCS filters are modular and independent of the network's architecture. This allows them to be easily deployed to image segmentation tasks. All networks were evaluated on their ability to segment the prostate gland from MRIs. Owing to shape variability and poor tissue contrast in patient MRIs, prostate boundary delineation is challenging. The OOCS networks were compared against their respective counterparts by performing a rigorous robustness analysis. 

\subsection{U-Net-Like Networks}
\label{Networks}
Along with the U-Net, we selected the Attention U-Net and the V-Net, as state-of-art U-Net benchmarks in the 3D medical image segmentation domain. These networks were selected based on the fact that they both introduced novel multi-view and multi-scale functions. The functions enhanced the receptive fields of the networks and reduced bottlenecks during computation, essentially shortening the time required to converge. All the standard networks were extended with 3D-OOCS filters of two different kernel sizes: $k=3$ and $k=5$, respectively. The new networks are called OOCS U-Net (k3), OOCS U-Net (k5), OOCS Attention U-Net (k3), OOCS Attention U-Net (k5), OOCS V-Net (k3), and OOCS V-Net (k5), respectively. 

\begin{table}[t]
\fontsize{9}{10}\selectfont
\centering
\begin{tabular}{ |c|c|c|c|}
\hline
    Parameter & Lower bound & Upper bound & Default value\\
    \hline
    $lr$  & $10^{-6}$ & 0.1 & $10^{-4}$  \\
    $\beta_1$ & 0.1 & 0.9 & 0.9  \\
    $dr$ & 0.0 & 0.7 & 0.5 \\
    \hline
\end{tabular}
\caption{Hyper-parameter ranges used in the optimal-configuration search for the baselines and their extension.}
\label{table:config_space}
\end{table}

\begin{table}[t]
\fontsize{9}{10}\selectfont
\centering
\begin{tabular}{|@{\ }c@{\ }|@{\ }c@{\ }|@{\ }c@{\ }|}
\hline                  
Model Name     & DSC & HSD(mm) \\
\hline
U-Net                     &  $0.744 \pm 0.24$ & $33.777\pm 37.81$ \\
OOCS U-Net (k3)           &  $0.798 \pm 0.11$ & $24.518\pm 12.46$\\
OOCS U-Net (k5)           &  $\mathbf{0.824 \pm 0.07}$ & $\mathbf{24.474\pm 12.48}$\\
\hline
V-Net                     &  $0.792 \pm 0.16$ & $25.170\pm 22.91$\\
OOCS V-Net (k3)           &  $0.791 \pm 0.13$ & $26.488\pm 17.68$\\
OOCS V-Net (k5)           &  $\mathbf{0.825 \pm 0.08}$ & $\mathbf{21.471\pm 10.01}$\\
\hline
Attention U-Net           &  $0.824 \pm 0.09$ & $27.822\pm 14.64$\\
OOCS Att. U-Net (k3)      &  $\mathbf{0.845 \pm 0.07}$ & $24.106\pm 14.70$\\
OOCS Att. U-Net (k5)      &  $0.835 \pm 0.11$ & $\mathbf{23.531\pm 14.89}$\\

\hline
\end{tabular}
\caption{Segmentation test results of different U-Nets and their OOCS extensions.}
\label{table:Performance-Table-Metrics-1}
\end{table}

\subsection{Dataset and Pre-Processing}
The Medical Segmentation Decathlon (MSD)\cite{MDC_2019} - prostate dataset comprises of 48 (training =32, testing =16) multimodal (T2, ADC) 3D MRI samples. However, the ground-truth labels for the testing images are not publicly accessible, and therefore, we implemented a 5-fold cross-validation. The available dataset (n=32) was split into training (80\%) and testing (20\%) samples for each fold (for details, please refer to the supplements file). Since the T2-weighted volumes provide most of the necessary anatomical information, and the ADC modality is often used for tumour characterization/segmentation, we only considered the T2-weighted modality for our experiments. Furthermore, the ground-truth labels of the original dataset are separated into annotations of the central prostate gland and the peripheral zone. We combined the two target regions into one such that the networks could learn binary image segmentations. The MRI scans and the ground truth labels were resampled to ($0.6$, $0.6$, $0.6$) voxel spacings with tri-linear and nearest-neighbor interpolation techniques, respectively. The resampling was done to increase the slice thickness and facilitate the smooth functioning of the networks without any alteration in their original architecture. 

\begin{table*}[t]
\fontsize{9}{10}\selectfont
\centering
\begin{tabular}{ |c|c|c|c| }
\hline                  
Model Name     & $\sigma=2$ & $\sigma=3$  & $\sigma=3.5$ \\
\hline
U-Net & $0.483 \pm 0.36$ & $0.242 \pm 0.33$ & $0.269 \pm 0.33$ \\
OOCS U-Net (k3) & $0.378 \pm 0.36$ & $0.311 \pm 0.37$ & $0.257 \pm 0.34$ \\
OOCS U-Net (k5) & $0.477 \pm 0.34$ & $0.344 \pm 0.39$ & $0.293 \pm 0.34$ \\
\hline
V-Net & $0.489 \pm 0.38$ & $0.243 \pm 0.35$ & $0.365 \pm 0.37$ \\
OOCS V-Net (k3) & $0.354 \pm 0.36$ & $0.312 \pm 0.39$ & $0.252 \pm 0.34$ \\
OOCS V-Net (k5) & $0.461 \pm 0.37$ & $0.341 \pm 0.40$ & $0.264 \pm 0.36$ \\
\hline
Attention U-Net & $0.688 \pm 0.23$ & $0.503 \pm 0.28$ & $\mathbf{0.533 \pm 0.28}$ \\
OOCS Att. U-Net (k3) & $0.691 \pm 0.16$ & $0.467 \pm 0.35$ & $0.427 \pm 0.34$ \\
OOCS Att. U-Net (k5) & $\mathbf{0.773 \pm 0.11}$ & $\mathbf{0.535 \pm 0.33}$ & $0.528 \pm 0.32$ \\
\hline
\end{tabular}
\caption{Robustness of U-Net variants against Gaussian Blur Noise (using DSC metric).}
\label{table:Performance-Table-Metrics-2}
\end{table*}

\begin{table*}[ht]
\fontsize{9}{10}\selectfont
\centering
\begin{tabular}{ |c|c|c|c| }
\hline                  
Model Name      & $\sigma=30$ & $\sigma=45$  & $\sigma=60$ \\
\hline
U-Net           & $0.737 \pm 0.23$ & $0.640 \pm 0.29$ & $0.458 \pm 0.38$ \\
OOCS U-Net (k3) & $0.792 \pm 0.12$ & $0.780 \pm 0.13$ & $0.746 \pm 0.18$ \\
OOCS U-Net (k5) & $0.822 \pm 0.07$ & $0.816 \pm 0.07$ & $0.807 \pm 0.08$ \\
\hline
V-Net           & $0.677 \pm 0.30$ & $0.633 \pm 0.30$ & $0.533 \pm 0.34$ \\
OOCS V-Net (k3) & $0.774 \pm 0.17$ & $0.714 \pm 0.24$ & $0.630 \pm 0.29$ \\
OOCS V-Net (k5) & $0.824 \pm 0.08$ & $0.822 \pm 0.08$ & $\mathbf{0.822 \pm 0.09}$ \\
\hline
Attention U-Net      & $0.795 \pm 0.15$          & $0.715 \pm 0.26$          & $0.614 \pm 0.32$ \\
OOCS Att. U-Net (k3) & $\mathbf{0.837 \pm 0.08}$ & $0.795 \pm 0.13$          & $0.697 \pm 0.25$ \\
OOCS Att. U-Net (k5) & $0.833 \pm 0.11$          & $\mathbf{0.827 \pm 0.12}$ & $0.813 \pm 0.13$ \\

\hline
\end{tabular}
\caption{Robustness of U-Net variants against Random Gaussian Noise (using DSC metric).}
\label{table:Performance-Table-Metrics-3}
\end{table*}

\begin{table*}[!htbp]
\fontsize{9}{10}\selectfont
\centering
\begin{tabular}{ |c|c|c|c| }
\hline                  
Model Name     & $transform\#=5$ & $transform\#=8$  & $transform\#=10$ \\
\hline
U-Net           & $0.629 \pm 0.26$          & $0.608 \pm 0.28$ & $0.564 \pm 0.30$ \\
OOCS U-Net (k3) & $0.484 \pm 0.24$          & $0.451 \pm 0.32$          & $0.506 \pm 0.31$ \\
OOCS U-Net (k5) & $0.668 \pm 0.10$ & $0.592 \pm 0.26$          & $0.613 \pm 0.24$ \\
\hline
V-Net           & $0.619 \pm 0.30$          & $0.674 \pm 0.26$ & $0.591 \pm 0.32$ \\
OOCS V-Net (k3) & $0.610 \pm 0.28$          & $0.613 \pm 0.29$          & $0.634 \pm 0.24$ \\
OOCS V-Net (k5) & $0.757 \pm 0.08$ & $0.669 \pm 0.26$          & $0.659 \pm 0.23$ \\
\hline
Attention U-Net      & $0.764 \pm 0.14$          & $0.806 \pm 0.09$          & $\mathbf{0.775 \pm 0.12}$ \\
OOCS Att. U-Net (k3) & $\mathbf{0.788 \pm 0.07}$ & $\mathbf{0.821 \pm 0.07}$ & $0.770 \pm 0.11$ \\
OOCS Att. U-Net (k5) & $0.771 \pm 0.10$          & $0.800 \pm 0.09$          & $0.747 \pm 0.15$ \\

\hline
\end{tabular}
\caption{Robustness of U-Net variants against Motion Noise (using DSC metric).}
\label{table:Performance-Table-Metrics-4}
\end{table*}

Furthermore, the 3D images were cropped during the experiments to a selected shape of $160\,{\times}\, 160\,{\times}\,64$ voxels per patch. For a given batch of samples, the intensity values of the MRIs were linearly scaled using a standard (z-score) normalization. Since the dataset is made up of a small number of volumes, the data pipeline carried out augmentations such as axial flips, elastic deformations, random crops, and affine transformations (scaling, rotation, and translation). 

\subsection{Implementation}
A combination of two-loss functions: Binary Cross-Entropy loss (L\textsubscript{BCE}) and Sørensen-Dice Loss (L\textsubscript{dice}) \cite{sudre_generalised_2017} called BCE-dice loss (L\textsubscript{BCE-dice}) was used to train the models. As demonstrated experimentally, this compound loss is defined over all semantic classes and is less sensitive to class imbalance \cite{MA2021102035}. The models were evaluated based on two metrics: DSC and symmetric Hausdorff Distance (HSD). They were chosen because they exhibit low overall biases in 3D medical image segmentation tasks \cite{taha_metrics_2015}. We used the Adam optimization algorithm along with a cosine-annealing learning rate scheduler \cite{adam_optimizer_2015, cosine_annealing_2017} throughout all the experiments. A batch normalization layer was placed after every convolution layer, to accelerate the speed of the experiments, and also serve as a regularizer~\cite{ioffe_batch_norm_2015}. All the networks were implemented in PyTorch~\cite{pytorch_2019_NIPS}. We ran our experiments without data-parallelism on the NVIDIA Titan RTX with 24 GB memory. The gradient updates were evaluated with a batch size of two samples.

\subsection{Training}
Individual network hyper-parameter searches were done with MRIs of fold-1. This ensured that all models achieved their peak performance, and therefore, a fair and objective comparison was conducted among all networks. The search space, as described in Table~\ref{table:config_space}, was used to fine-tune the learning rate ($lr$), Adam's momentum ($\beta_1$), and the dropout rate ($dr$). Please see the supplements for the table of the tuned parameters found for each network. After finding the optimal hyper-parameters for a network, four additional models of the same network were trained using the four remaining cross-validation folds, and the fine-tuned values of $lr$, $\beta_1$ and $dr$. This kind of training was repeated for all  the networks. 

\subsection{Results}
\subsubsection{Testing}
The testing of the networks was done employing the sliding window approach where the window size equals the patch size used during training. The average DSC and HSD of all the networks are listed in the Table~\ref{table:Performance-Table-Metrics-1}. Here, it is evident that the OOCS-Attention-U-Net variations performed best. In addition, the OOCS extensions have consistently outperformed their original counterparts in most instances. The superior performance of the OOCS models showcases the ability of the inductive bias in the OOCS filters to delineate prostate contours effectively. 

\subsubsection{Robustness Evaluation}
To examine the robustness of the models, we introduced three types of noise: Gaussian blur, random Gaussian noise, and motion transform (simulating artifacts due to patient movement during acquisition), into the test volumes. In case of blurring, the average DSCs of the models decreased with the increase in standard deviation ($\sigma$) of the noise level. This behaviour was anticipated since higher deviations produced samples with large perturbations. On the other hand, the prostate predictions of the OOCS models were considerably stable across both random Gaussian and motion artifacts. This demonstrates that the OOCS-extensions have a lower variance when compared to the benchmark networks on complex test sets and in the presence of  outliers. Finally, the OOCS-Attention-U-Net (k5) significantly outperforms all U-Net-like benchmarks and their extensions.
Tables~\ref{table:Performance-Table-Metrics-2},~\ref{table:Performance-Table-Metrics-3} and ~\ref{table:Performance-Table-Metrics-4} summarise the results of the  robustness evaluation of the OOCS models and the state-of-the-art models. Please refer to the supplement for the illustration of a few of the examples with noise information, along with the predictions of the benchmark and the OOCS Attention U-Net (k5) models, respectively.

\section{Conclusion}
We expanded the On-Off center-surround retinal receptive fields to 3D kernels, and proposed a straightforward method to calculate the kernel weights. We designed OOCS encoder blocks using these fixed and pre-computed kernels as edge detection inductive biases, with On and Off parallel pathways similar to the ganglion pathways in the visual system. Our proposed 3D OOCS encoder block does not increase the number of training parameters of the networks and requires minimal computational overhead. Moreover, it is effortless to equip any UNet-like network with these encoder blocks.
Our experiments on prostate segmentation tasks with our baseline networks and their OOCS extensions show notable enhancements in the performance of the OOCS-equipped networks, thus making 3D-OOCS a favorable method to achieve accurate results in 3D medical image segmentation tasks. Furthermore, our robustness experiments demonstrate the superior performance of OOCS extended networks. Robustness to distribution shifts is a crucial attribute of any deep neural network, and it is of even higher importance for networks used in critical domains like bio-medical tasks.

\section{Acknowledgements}
This research was funded in part by the Austrian Science Fund (FWF): I 4718, and the Federal Ministry of Education and Research (BMBF) Germany, under the frame of ERA PerMed. Z. B. was supported by the Doctoral College Resilient Embedded Systems, which is run jointly by the TU Wien's Faculty of Informatics and the UAS Technikum Wien.


\section{Code and Data Availability} 
All code and data are included in https://github.com/Shrajan/AAAI-2022.

\bibliography{aaai22}

\end{document}